\documentclass[prb,aps,twocolumn,floats]{revtex4}
\usepackage{amstext}
\usepackage{amsmath}
\usepackage{graphics}
\usepackage{multicol}
\usepackage{epsfig}

\def\asinh{{\rm asinh}}

\begin{document}

\title{{\em Ab initio} and finite-temperature molecular dynamics
studies of lattice resistance in tantalum}
\author{D.E. Segall\footnote{ Current Address: Department of Applied
Physics, California Institute of Technology, Pasadena, California
91125.}}  \address{Department of Physics, Massachusetts Institute of
Technology, Cambridge MA 02139}

\author{Alejandro Strachan}\address{  Materials and
Process Simulation Center, Beckman Institute (139-74)  California Institute of Technology, Pasadena, CA 91125}

\author{Sohrab Ismail-Beigi} \address{Department of Physics, University of California at Berkeley,
Berkeley, CA 94720}

\author{William A. Goddard~III}\address{  Materials and
Process Simulation Center, Beckman Institute (139-74)  California Institute of Technology, Pasadena, CA 91125}
\author{T.A. Arias}  \address{Laboratory of Atomic and Solid State Physics, Cornell University,
Ithaca, NY 14853}

\begin{abstract}
This manuscript explores the apparent discrepancy between experimental
data and theoretical calculations of the lattice resistance of bcc
tantalum.  We present the first results for the temperature dependence
of the Peierls stress in this system and the first {\em ab initio}
calculation of the zero-temperature Peierls stress to employ periodic
boundary conditions, which are those best suited to the study of
metallic systems at the electron-structure level.  Our {\em ab initio}
value for the Peierls stress is over five times larger than current
extrapolations of experimental lattice resistance to zero-temperature.
Although we do find that the common techniques for such extrapolation
indeed tend to underestimate the zero-temperature limit, the amount of
the underestimation which we observe is only 10-20\%, leaving open the
possibility that mechanisms other than the simple Peierls stress are
important in controlling the process of low temperature slip.
\end{abstract}

\twocolumn

\maketitle

\section{Introduction} \label{sec-intro}

The study of the plasticity of crystalline materials is a rich
many-body problem involving physics on multiple length scales, with
many remaining unexplained mysteries.  The plasticity of bcc metals,
for instance, is particularly challenging.  Unlike their fcc and hcp
counterparts, the bcc metals exhibit many active slip planes, have a
strong temperature dependence in their plasticity and violate the
simple empirical Schmid law~\cite{schmid}.  Moreover, theoretical
calculations of the most basic question in plasticity, the stress
needed to induce yield at low temperature in a pure sample, differ
from experimental extrapolations by over a factor of
two~\cite{duesbery3}.  The purpose of this work is to provide needed
insight into this discrepancy.

Ultimately, it is the physics of the $\langle 111\rangle $ screw
dislocation defect which controls the low-temperature plasticity of
bcc materials~\cite{vitek,duesbery3}.  The Peierls stress, the yield
stress at which these dislocations first begin to move spontaneously,
is difficult to compare directly with experiment.  Whereas most
computational work on the Peierls stress measures the stress to move
an isolated, infinitely straight dislocation at zero
temperature~\cite{vitek,duesbery1,segall,wang,rao1,woodward1,woodward2,yang,xu,moriarty,mgpt,takeuchi,ito},
experiments measure the Peierls stress at a finite temperature in
systems with many interacting, curved dislocations and in media with
defects and surfaces.  As an example of the present challenges which
the literature faces, using model generalized pseudopotential theory
(MGPT)~\cite{mgpt}, Yang {\em et. al.}\cite{yang} predict for the T=0
Peierls stress a value 2.5 times greater than experimental
extrapolations~\cite{tang}.  Because such potentials are not based
upon first principles, it is impossible to determine {\em a priori}
whether this discrepancy is due to the inter-atomic potential, the
environmental complexities discussed above, or to a flaw in our
understanding of the relation between the Peierls stress and the
experiments.

As it is a daunting experimental task~\cite{sigle} to observe
properties of a single dislocation locked deep in the heart of a
material, accurate theoretical calculations of such systems is
essential.  Nearly all theoretical calculations to date, concerning
such dislocations, have relied upon empirical
potentials~\cite{vitek,duesbery1,segall,wang,shenoy,moriarty,xu,mgpt,ito,yang,rao1,bulatov,takeuchi}.
Given the empirical nature of such calculations, the complex
directional bonding properties of bcc materials, and the lack of
direct comparison with experiments for validation, first principles
{\em ab initio} calculations of dislocations in such systems are
clearly needed.  Ismail-Beigi and Arias\cite{sohrab_disloc} were the
first to show that density functional theory calculations were crucial
in understanding the fundamental properties of the $\langle
111\rangle$ screw dislocation core structure in bcc molybdenum and
tantalum.  Until that work, most computational studies based on
empirical
potentials~\cite{vitek,duesbery1,rao1,mgpt,wang,xu,moriarty,segall},
supported the idea that the dislocation core breaks symmetry, with two
energetically equivalent ground state structures which spread outward
along two different equivalent sets of three $\{110\}$
planes\cite{vitek}, similar to the concept originally proposed by
Hirsch~\cite{hirsch,mitchell}.  Until the availability of the {\em ab
initio} calculations, the prevailing view of the violation of the
Schmid law in the bcc metals was based upon this
structure~\cite{vitek}.  Ismail-Beigi and Arias~\cite{sohrab_disloc},
in contrast, showed that for both molybdenum and tantalum the
ground-state structure within density functional theory was a
non-degenerate symmetric core, strongly supporting the work of Suzuki
and Takeuchi~\cite{suzuki,takeuchi,kimura} which first suggested that
it is the Peierls potential itself that controls the lattice
resistance and not the details of the core structure.  To help resolve
the discrepancy between theoretical and experimental Peierls stresses,
the work below provides a reliable {\em ab initio} prediction of the
Peierls stress in bcc tantalum which is free of the unrealistic
electronic boundary conditions employed in the only other {\em ab
initio} prediction of the Peierls stress\cite{woodward1,woodward2}.

Here, we show that the Peierls stress, calculated within density
function theory, is over a factor of {\em five} larger than expected
from extrapolation of experimental results~\cite{tang}.  This supports
the view that the discrepancy between the experimental and
computational predictions are largely due to the aforementioned
environmental complexities, to a flaw in relating the experimental
data to the Peierls stress, or to a combination of both.

To further explore possible physical effects leading to this
discrepancy, we study the extrapolation of experimental data to
determine the zero-temperature Peierls stress.  Such extrapolations
generally employ fits from mesoscopic or thermodynamics/kinetic
models~\cite{tang,kocks,cuitino,stainer}.  However, it has not been
established that such models can accurately describe the lowest
temperature regime correctly, placing doubt on the quality of these
extrapolations.  To address this issue, the work below also provides
the first temperature- and orientation- dependent study of the Peierls
stress in a bcc metal.  We moreover show that extrapolation of our
finite temperature results using a current fitting model leads to an
underestimation of the zero-temperature Peierls stress.  This
underscores the difficulty in extrapolating the experimental data
accurately but does not fully account for the observed discrepancy.

In Section~\ref{sec-bc}, this manuscript reviews the various
techniques in use for calculation of the Peierls stress in the context
of efficacy for application to {\em ab initio} calculations.
Section~\ref{sec-tempdep} gives the first calculation of the
temperature and orientation dependent Peierls stress in a bcc
material.  Section~\ref{sec-conv} describes our technique for
obtaining Peierls stresses within small unit cells with periodic
boundary conditions.  Finally, Section~\ref{sec-results} presents our
{\em ab initio} prediction for the Peierls stress and compares and
contrasts it to currently available experimental and computational
values.

\section{Boundary Conditions}\label{sec-bc}

The fundamental distinction among theoretical approaches to
calculation of the Peierls stress is the choice of boundary condition.
The literature describes three types of boundary conditions:
cylindrical boundary
conditions~\cite{vitek,segall,shenoy,mgpt,xu,takeuchi,ito,duesbery1},
Greens function (or ``flexible'') boundary
conditions~\cite{sinclair,rao2,yang,rao1,vitek,woodward1,woodward2},
and periodic boundary conditions~\cite{segall,wang,bulatov}.  We now
briefly review each with emphasis on the unique challenges of {\em ab
initio} electronic structure calculations.

\subsection{Cylindrical Boundary Conditions}

In the practice of cylindrical boundary conditions, anisotropic
elasticity theory~\cite{hirth,stroh,head} is used to generate a
dislocation in the center of a cylinder. The cylinder is then
separated into {\em inner} and {\em outer} regions.  The atoms in the
outer region are held fixed to the solution of anisotropic elasticity
theory while the atoms in the inner region relax under the
inter-atomic forces. To calculate the Peierls stress, a stress is
applied to the system until the dislocation moves.

This approach suffers numerous draw backs when applied to density
functional theory. To avoid surface effects and to properly account
for the non-linear nature of the dislocation, such cylinders generally
have to be quite large.  First, even the outer cylinder is of finite
size and therefore the outer region must be sufficiently large enough
so that forces generated by it onto the inner region are equivalent to
those generated from an infinite continuum. The inner region also must
be sufficiently large to mitigate two effects.  The inner region must
be large enough so that linear elasticity theory represents well the
forces which it imposes on the outer region.  The inner region also
must be large enough so that motion of the dislocation is not
adversely affected by the fixed outer region, which is a concern
because the fixed outer region represents the displacement field when
a dislocation is at its center and therefore generates a extraneous
force which tends to prevent motion of the dislocation~\cite{shenoy}.
When using simple, inter-atomic potentials, the use of large cylinders
mitigates all of these effects.  However, this approach is not viable
for density functional calculations with their extreme computational
demands.

This approach, moreover, is particularly ill-suited for electronic
structure calculations because the artificial surface at the outside
of the outer region, being far different from the bulk, give rise to
strong scattering of the electrons far different than would an
infinite continuum.  This is particularly problematic for metals,
because the single-particle density matrix, which quantifies the
effects of this scattering on the inter-atomic forces, decays only
algebraically in metals~\cite{sohrab_densmat}.  The following
subsection demonstrates that the boundary regions should be quite
large in order for these surface effects not to result in large
fictitious forces in the active region of the calculation.

\subsection{Greens Function Boundary Conditions}

The use of Greens function, or flexible, boundary
conditions~\cite{sinclair,rao2,vitek}, is an effective way to reduce
the size of the simulation cell.  This approach also employs a
cylindrical geometry.  However, rather than the ``inner'' and
``outer'' atomistic regions of the cylindrical boundary approach, the
Greens function approach employs three inter-atomic regions: an inner
``core'' region containing the center of the dislocation, an
intermediate ``buffer'' region, and an outermost
``continuum-response'' region.  With {\em proper} implementation, the
outer and inner regions couple only indirectly through the response of
the buffer region.

In this method, all three regions respond to the presence of a
dislocation; however the response of each region is treated
differently through a number of steps.  Initially, all regions are
displaced by the solution to anisotropic elasticity theory.  Each
iteration then begins by relaxing the atoms in the core region
according to the forces which they experience, as computed from either
an inter-atomic potential or an {\em ab initio} method.  The forces
generated from the mismatch between the outer and inner regions, which
the cylindrical approach above ignores, are then relieved by moving
the atoms of {\em all three regions} according to the elastic Greens
function solution, leaving only the nonlinear effects from the core
region unaccounted.  The next iteration then begins by relaxing these
forces as described above.  Iterations proceed until until the forces
in the core and buffer region are negligible.

What distinguishes this approach from simple cylindrical boundary
conditions is that the continuum region, via the Greens function
response, is allowed to respond to the motion of the dislocation and
to the elastic response generated by the core region as the
dislocation moves.  So long as the continuum response region (a)
accurately represents the structure induced by the presence of the
dislocation and (b) is sufficiently wide to properly reproduce the
forces on the atoms in the buffer and inner regions, this approach
accurately describe basic properties of a dislocation.

In order for the first assumption (a) above to hold, the inner core
region must be sufficiently large to contain all atoms with
displacements outside of the linear regime and the buffer region must
be sufficiently wide so that displaced atoms in the core have no
effect on the forces experienced in the continuum-response region.
The second assumption (b) requires that the continuum-response region
to be sufficiently large so that its termination has no effect on the
forces on the atoms in the buffer or inner region.  The radius of the
calculation must therefore exceed the sum of the non-linear core
radius plus twice the range over which motion of atoms creates forces
within the lattice.  As the latter range can be quite large for
electronic structure calculations in metals, the application of this
approach to electronic structure calculations can be problematic.

The Greens function approach has predicted successfully dislocation
properties when applied to time consuming empirical
potentials~\cite{rao1,yang} which have a limited interaction range.
The approach also has been applied to density functional calculations
of the Peierls stress for molybdenum and
tantalum\cite{woodward1,woodward2}, where its application is more
questionable due to the above interactions.  In these latter works,
the artificial boundary on the outside of the continuum-response
region have been treated in one of two
ways~\cite{woodward1,woodward2}, either by keeping the surface free in
vacuum or by embedding in periodic boundary conditions with the vacuum
filled with material which must contain severe domain boundaries due
to the incompatibility of a net Burgers vector with periodic boundary
conditions.

To gauge the effects which this artificial boundary may have and how
far these effects penetrate from the continuum-response region into the
buffer region, we perform a test calculation within the density
functional theory pseudopotential approach~\cite{payne} of the
magnitude of the forces generated onto the system due to the presence
of a domain boundary similar to those in the works cited
above~\cite{woodward1,woodward2}.

For this calculation, we employ the same computational procedure as
for our production calculations in Section~\ref{sec-results}.  Here,
however, as this is a test, we employ only a single $k$-point to
sample the Brillouin zone ($\Gamma$).  We begin with an orthorhombic
cell of 24 atoms of tantalum in a bulk arrangement with supercell
lattice vectors $\vec r_1 = a[ 1 \bar 1 0 ]$, $\vec r_2 = 4 a[ 1 1
\bar 2 ]$ and $\vec r_3 = a/2 [ 1 1 1 ]$.  We choose this cell because
its length along $\vec r_2$ is the same as the smallest simulation
cell used in
References~\onlinecite{woodward1}~and~\onlinecite{woodward2}.  We then
generate a domain boundary at the edge of cell along the $(11\bar2)$
plane by changing the lattice vector $\vec r_2$ to $\vec r_2 = 4 a[ 1
1 \bar 2 ] + \alpha \vec r_3$ and holding the atoms in the unit cell
fixed in their bulk locations. $\alpha$ is chosen such that the shift
is small and the nearest neighbor distance is always within $95\%$ of
the bulk, representing even less of a disturbance that in
Reference~\onlinecite{woodward1}, where atom were within $90\%$ of the bulk
nearest neighbor distance.  To estimate the effect of the scattering
of electrons at the domain boundary on the inter-atomic forces, we
hold the atoms fixed and compute the {\em ab initio} forces acting
upon them.

\begin{figure}
\begin{center}
\rotatebox{90}{\hspace*{.2in} Force along [111]
[eV/\AA]} \scalebox{0.35}{ \includegraphics{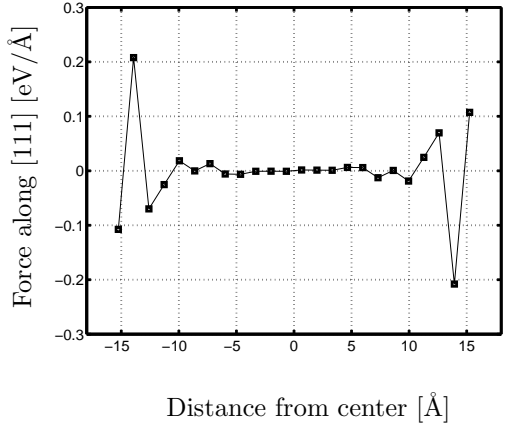}} \\ \ \\
\hspace*{0.5in} {Distance from center [\AA]}
\end{center}
\caption{Force on the atoms, along the $[ 111 ] $, due to
the presence of a domain boundary. The forces are plotted as a
function of distance from the center of the unit cell. The domain
boundary is generated such that the nearest neighbor distance is
always within $95\%$ of the bulk ($\alpha$ = 1/4).}
\label{fig:domain}
\end{figure}

Figure~\ref{fig:domain} shows the forces along the $[ 111 ]$ direction
as a function of distance from the center of each domain.  Note that
relatively large forces develop deep within the cell.  This data
indicates that the continuum-response region should be quite large
($\approx 5-10$~\AA) in order for the response of the electrons not to
adversely effect the forces in the buffer region.  Note also that the
the buffer region should be of similar width to prevent forces from
the non-linear displacements in the core from penetrating into the
linear continuum-response region.  Such large continuum-response and
buffer regions can make the calculation infeasible with current
computational techniques.

In fact, the only density functional calculations of the Peierls
stress in this system to date employ the Greens function method but
with a distance from the buffer region to the domain boundary of only
$\approx 3.7$~\AA.  It thus is unclear whether the continuum region in
these calculations is sufficiently large to lead to reliable results
and clearly further calculations are needed to support those results.
Below, we provide just such calculations using the method of periodic
boundary conditions, which perturb the electronic system far less than
the introduction of domain boundaries.

\subsection{Periodic Boundary Conditions} \label{sec:PBCs}

The final common choice for boundary conditions is to repeat the
dislocation core periodically throughout space, so that the
dislocation is no longer isolated, but embedded in bulk material
containing an array of dislocations.  Consistency with periodic
boundary conditions then demands that the unit cell contain an even
number of dislocations with Burgers vectors of alternating sign,
arranged typically in dipolar~\cite{marklund,ariasjoannop,lehto} or
quadrupolar~\cite{bigger,lehto,sohrab_disloc} arrays.  For static
properties of the dislocation core, such cells give reliable results
as the elastic fields of the surrounding dislocations effectively
cancel at the location of each core.

Periodic boundary conditions can also been used to calculate the
Peierls stress~\cite{segall,wang,bulatov}. Care must be taken,
however, because the symmetry of the dipole/quadrupole array breaks as
the dislocations move and the dislocation-dislocation interaction then
ceases to be negligible.  The use of large unit cells can control this
effect\cite{bulatov}; however, such a direct, brute-force approach is
not practical for computationally demanding {\em ab initio}
calculations.  To make such calculations feasible, they must occur in
small periodic cells, thereby demanding proper accounting of the
dislocation-dislocation interactions.

We have shown in another work~\cite{segall} that, under certain
conditions, such interactions can be accounted accurately with minimal
extra computational effort, so that accurate values of the Peierls
stress can be obtained from density functional calculations in
periodic cells.  As the residual errors with this approach are
associated with the boundary conditions, the magnitude of such an
error can be tested by other computational methods, such as empirical
potentials.  Our previous work shows that that this residual error is
relatively small~\cite{segall}, a fact which we confirm explicitly
below.

Because the deviations from the bulk arrangement at periodic
boundaries are relatively mild, such calculations are ideal for
mitigating electronic boundary effects.  Given the simplicity of
working with these boundary conditions and the possibility of the
extraction of accurate values for the Peierls stress from small unit
cells, we choose to work with periodic boundary conditions throughout
this work.  Section~\ref{sec-conv} outlines our procedure for
calculating the Peierls stress while working with periodic boundary
conditions and describes the sources and the magnitude of the residual
errors.  (See Reference~\onlinecite{segall} for a full discussion of
these issues.)

\section{Dependence of the Peierls stress on orientation and
temperature}\label{sec-tempdep}

To illustrate the complexities of relating computational predictions
to experimental findings, we now explore the dependence of the Peierls
stress in bcc tantalum on orientation and temperature.  The strong
dependencies which we shall find underscore the unique properties of
dislocations in bcc metals. To clarify, as some authors use slightly
different definitions for the Peierls stress, here we consider the
Peierls stress as the value of the stress on the maximum resolved
shear stress plane (maximum value of the shear stress along the
[111]-direction) when the dislocation {\em first} moves to a different
equilibrium position.

Despite recent advances in {\em ab initio} quantum mechanical methods,
such methods are still too computationally intensive to study such
properties as the temperature dependence of the Peierls stress.
Therefore, for these calculations, we employ a molecular dynamics (MD)
framework carried out using a first-principles-based, many body force
field (FF) for tantalum, which we denote qEAM, which we have developed
to allow accurate and computationally efficient evaluation of atomic
interactions~\cite{strachan,wang}.

As described above, we carry out these calculations within periodic
boundary conditions.  The super-cell consists of a quadrupolar
arrangement~\cite{bigger,lehto} of dislocations containing 5670 atoms
with lattice parameters $a_x = 70.59$~\AA, $a_y = 73.39$~\AA \ and
$a_z=20.11$~\AA, where the $x$-, $y$-, and $z$- axes of our coordinate
system are along $[ 1\bar 1 0 ]$, $[ 11\bar 2 ]$, and $[ 111 ]$
directions, respectively.  As we have shown in another
work~\cite{segall}, such a cell gives very accurate values for the
Peierls stress.

\subsection{Orientation dependence of the zero-temperature Peierls
  stress of the $\langle 111 \rangle$ screw dislocation}

To calculate the zero temperature Peierls stress, we start with a
fully relaxed quadrupole dislocation configuration at zero stress and
increase the stress in steps of 50 MPa until the dislocations move.
Once the dislocations move, we restart the calculation from the
structure equilibrated just prior to the motion and increase the
stress in smaller steps (5 MPa) in order to more narrowly define the
critical stress.  At each incremental target stress, we relax the
atoms and stresses in the cell by running two very low temperature
(T=0.001 K) MD simulations.  The first run is for 15 ps at constant
stress and temperature (N$\sigma$T ensemble) using a Rahman-Parrinello
barostat~\cite{rahman} and a Hoover~\cite{hoover} thermostat, and the
second run is for 50 ps at constant volume and temperature (NVT
ensemble).  We find this approach to be quite stable for relaxing the
cell and the atoms of the system.

The $\langle 111 \rangle$ screw dislocation has three equivalent
\{112\} and three equivalent \{110\} potential slip planes, with such
planes occurring at 30$^\circ$ intervals.  To study the orientation
dependence of the Peierls stress, we apply three types of pure shear
stress to the system: a $\sigma_{xz}$ stress, a positive $\sigma_{yz}$
stress and a negative $\sigma_{yz}$ stress.  (Note that with
coordinate axes as defined above, the $z$-axis lies along the
dislocation line.)  These stresses leads to forces on the dislocation
in the $\langle 112 \rangle$, $\langle 110 \rangle$-twinning and
$\langle 110 \rangle$-antitwinning directions,
respectively\cite{hirth}.  Along these directions, we find Peierls
stresses of $\tau_{112}$~=~655~MPa, $\tau_{twin}$~=~575~MPa, and
$\tau_{antitwin}$~=~1075~MPa, respectively.
Table~\ref{table_peierls_orint} shows that our essentially
zero-temperature results are in good agreement with those of Yang and
collaborators~\cite{yang}, who employed model generalized
pseudopotential theory (MGPT)\cite{mgpt}, a different inter-atomic
potential.  The result of such a strong dependence of the Peierls
stress on orientation is consistent with the experimentally observed
breakdown of the Schmid law in bcc metals.

\begin{table}
\begin{tabular}{|l|c|c|c|c|}
\hline
 Potential& Twin & $\langle 112 \rangle $ & Anti-Twin & Asymmetry \\
\hline
qEAM FF       &   575 MPa   &   655 MPa  & 1075 MPa & 1.6412 \\
MGPT \cite{yang}   &   605 MPa   &   640 MPa  & 1400 MPa & 2.29   \\
\hline
\end{tabular}
\caption{Peierls stress for the $\langle 111 \rangle$ screw
dislocation in tantalum in the twinning, $\langle 112 \rangle$ and
anti-twinning directions; the last column shows the ratio between
anti-twinning and twinning Peierls stresses.  MGPT results from Yang
et al.\cite{yang}}
\label{table_peierls_orint}
\end{table}

To make quantitative comparison with experiments, which are carried
out at nonzero temperature, we compare our results to those of Tang
{\em et. al.}~\cite{tang}, who fit experimental data~\cite{wasserbach}
to a mesoscopic model and then extrapolate to extract the
zero-temperature Peierls stress. Their predicted value of 248~MPa for
the $\langle 112\rangle$ Peierls stress is over a factor of two lower
than our result.  
%Although Tang and collaborates~\cite{tang} assumed
%the slip planes to be \{110\}, while the motion here was resolved 
%along a {112}-plane
This type of discrepancy, where the theoretical Peierls stress
overestimates the zero-temperature extrapolation of the experimental
data by a factor of two to three, is quite generally
observed\cite{duesbery3}.  This discrepancy may be due either to
inaccuracies in the theoretical calculations or, perhaps, to a flaw in
the comparison between the zero-temperature extrapolation of the
experimental data and theoretical predictions.

\subsection{Temperature dependence of the Peierls stress of the
$\langle 111 \rangle $ screw dislocation}

To explore potential difficulties with the zero-temperature
extrapolation, we now present what to our knowledge is the first
temperature-dependent study of Peierls stress using a realistic
potential for a bcc metal.  For these calculations, we continue to
employ the qEAM FF and begin with the zero-temperature, equilibrated
structures.  We then apply various constant shear stresses (lower than
the T=0 Peierls stress) to the system while slowly increasing the
temperature (in steps of 10~K) until the dislocations move.  Similarly
to the T = 0.001K case, for each temperature we first run for 10 ps
in the N$\sigma$T ensemble and then for 25 ps in NVT ensemble.

Because the Peierls stress can depend on the rate at which the strain
is applied, to place our results in context, we first estimate the
strain rate in our computations.  The strain rate is approximately
$\dot{\gamma} = \rho v_d b$, where $v_d$ is the dislocation velocity,
$\rho$ is the dislocation density and $b$ is the Burgers vector.
Using a dislocation density typical of the experiments\cite{tang}
($\rho =10^{11} 1/m^2$) and estimating the dislocation velocity as the
ratio between the distance traveled in one jump ($1/3 a\langle
112\rangle = 2.717$~\AA) and the simulation time (35 ps), we obtain an
effective strain rate of $\sim 10^2 1/s$, which is large compared to
the strain rates ($4\times 10^{-5}$) in the experiments used for
the zero-temperature extrapolations\cite{tang,wasserbach}.

\begin{figure}
\begin{center}
\scalebox{0.35}{\includegraphics{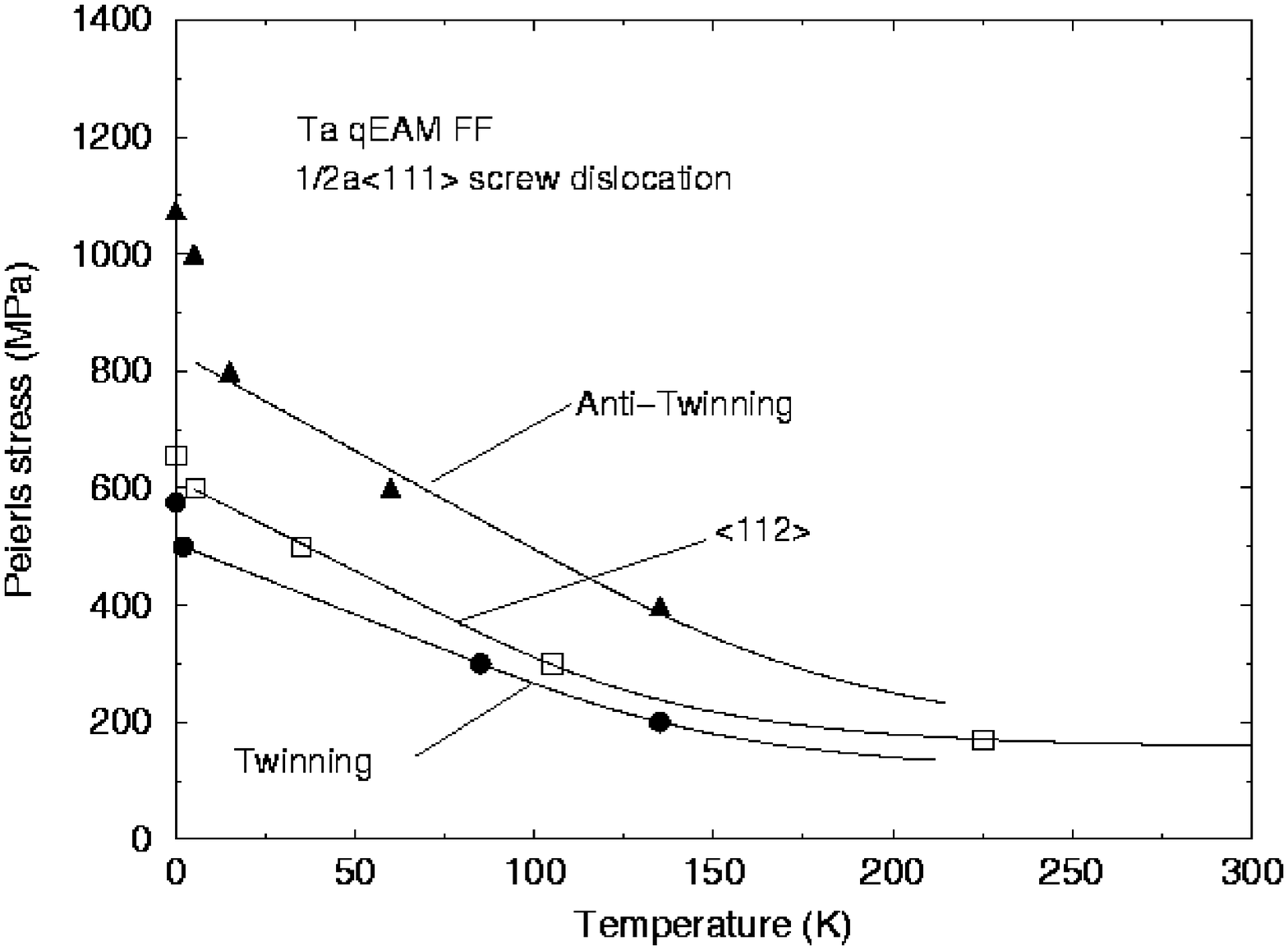}}
\end{center}
\caption{Temperature dependence of the Peierls stress along various
directions: the $\langle 112 \rangle$, twinning and anti-twinning
directions. The fits are done for the high temperature data using
Equation~(\ref{eqn:YieldStressDependence}). The temperature is in
Kelvin, and the stress in MegaPascals.}
\label{fig:tempdep}
\end{figure}

Figure~\ref{fig:tempdep} summarizes our results for the temperature
dependence of the Peierls stress as a function of temperature for the
three directions ($\langle 112 \rangle$, twinning and anti-twinning).
As expected, the Peierls stress obtained from our MD simulations
decreases rapidly with increasing temperature, particularly for very
low temperatures.  It is important to mention that, although our
simulations are three dimensional, the dislocations move as straight
lines without the formation of double kinks because our simulation
cell is only seven Burger's vectors long along the dislocation
lines. Such double kinks are quite important at finite temperatures as
they tend lower the lattice resistance at non-zero temperatures.  Our
results are approximately a factor of $2-4$ larger than the fit of
Tang {\em et.  al.}~\cite{tang} to the experimental data of
Wasserb\"{a}ch~\cite{wasserbach}.  We feel that this is reasonable,
considering the facts that our simulation cells do not allow for
double kink formation and that, as discussed above, our strain rates
are much higher than those in the experiments\cite{tang,wasserbach}.

Experimental extrapolations of the Peierls stress to zero temperature
generally come from mesoscopic or kinetic/thermodynamic
model~\cite{tang,kocks,cuitino,stainer} fits to experimental data and
extrapolated to zero temperature.  To explore the effects of this
procedure, we fit our atomistic data to such a model, perform the
extrapolation and then compare with our direct zero-temperature
results.

For the fit we use an analytical expression for the dependence of the
Peierls stress with temperature at constant strain
rate~\cite{stainer,cuitino} based on a mechanism involving double kink
nucleation and propagation. This model gives for the temperature
dependent Peierls stress
\begin{equation}
\tau_P = \frac{\tau_0}{\beta E^{\rm kink}}
\asinh \left(\frac{\dot{\gamma}}{\dot{\gamma}^{\rm kink}_0}{\rm
e}^{\beta E^{\rm kink}} \right),
\label{eqn:YieldStressDependence}
\end{equation}
where $\beta$ is $1/k_B T$, with $k_B$ being Boltzmann's constant and $T$
the absolute temperature, $E^{\rm kink}$ is the kink
energy~\cite{hirth}, $\tau_0$ is the effective Peierls stress and
$\dot{\gamma}^{\rm kink}_0$ is the reference strain rate.  Here, the
effective Peierls stress is
\begin{equation}
\label{eqn:tauzero}
\tau_0 = \frac{E^{\rm kink}}{b L^{\rm kink} l_P}
\end{equation}
and the reference strain rate is
\begin{equation}
\label{eqn:gammakink}
\dot{\gamma}^{\rm kink}_0 = 2 b \rho l_P \nu_D,
\end{equation}
where $b$ is the Burgers vector, $L^{\rm kink}$ is the kink length,
$\rho$ is the dislocation density and $\nu_D$ is the attempt frequency
which may be identified with the Debye frequency to a first
approximation~\cite{stainer,cuitino}, and $l_P$ is the distance
between two consecutive Peierls valleys.  Physically, $E^{\rm kink}$
is the minimum energy to form a double kink, $L^{\rm kink}$ is the
minimum length for this double kink and $\tau_o$ is the stress, whose
work to move a dislocation a distance $l_P$ is equal to $E^{\rm
kink}$.

Figure~\ref{fig:tempdep} shows the fit of
Equation~(\ref{eqn:YieldStressDependence}) to our atomistic data.  To
mimic how zero-temperature lattice resistances are generally
extracted, we adjust the three unknown parameters ($\tau_0, \ E^{\rm
kink} \mbox{ and } \dot{\gamma}^{\rm kink}_0$) to fit our higher
temperature data {\em only}.  Intriguingly, extrapolation of our
higher temperature results to zero-temperature leads to an
underestimation of the Peierls stress of between $10\%$ and $20\%$.
We would also expect that a fit to data from a cell sufficiently large
to allow for double kink formation (which is active in the experiments
at non-zero temperatures) would lead to an even a larger
underestimation of the zero-temperature stress.  These results
therefore suggest that the general discrepancy between extrapolated
experimental values and the calculated values for the zero-temperature
Peierls stress may be the result of failure of nonzero-temperature
models to describe properly the low-temperature regime.  This
illustrates one possible difficulty in relating the experiments to
computational predictions and underscores the need for first
principles studies of such a phenomena.

\section{Accurate Peierls Stress Calculations in small periodic cells}\label{sec-conv}

Having underscored the need for first principles electronic structure
studies and already determined the most effective boundary conditions
for such studies as periodic, we now focus on determination of the
cell of minimal size appropriate to calculation of the
zero-temperature Peierls stress for a $\langle 111 \rangle $ screw
dislocation in a bcc metal when the maximum resolved shear stress 
is along a \{110\} plane.

\begin{figure}
\begin{center}
\rotatebox{90}{\hspace*{0.8in}{\Large  $[  1 1 \bar 2 ]
\rightarrow $}}
\scalebox{0.4}{   \includegraphics{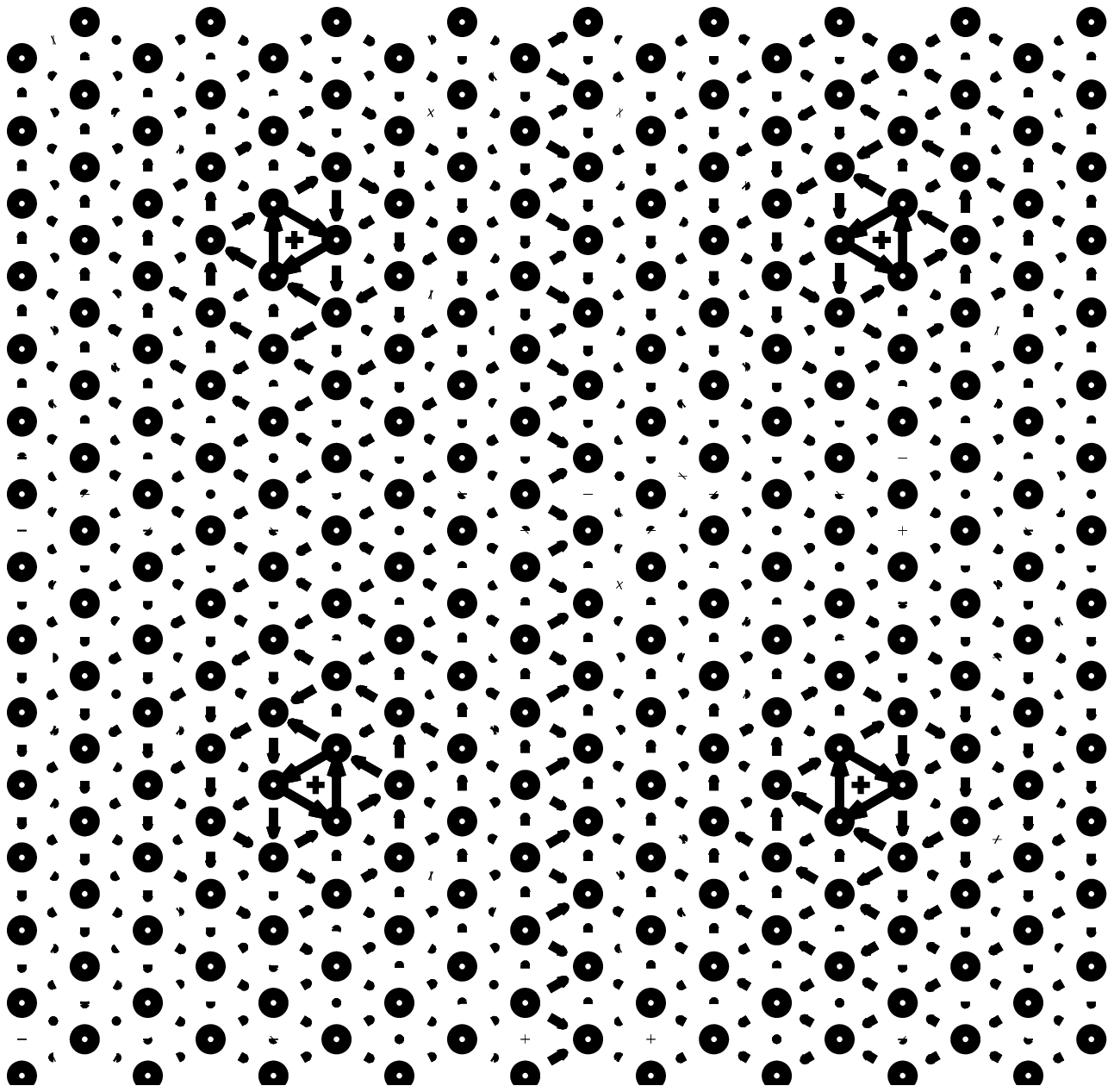}} \\ \ \\
\hspace*{0.5in} {\Large $[  1 \bar 1 0 ] \rightarrow $}
\end{center}
\caption{Quadrupolar unit cell of size 42\AA $\times$ 41\AA: relative
displacements of neighboring columns of atoms (arrows), dislocation
centers (plus signs).}
\label{fig:quad}
\end{figure}

To minimize image effects, we employ periodic boundary conditions with
a quadrupolar unit cell.  Figure~\ref{fig:quad} shows a differential
displacement map~\cite{vitek} of such a cell of size 42\AA $\times$
41\AA \ in the plane perpendicular to the Burgers vector.  In such a
map, the dots indicate columns of atoms along the $[ 111 ]$, and the
vectors between the columns of atoms indicate the relative shift along
the Burgers vector due to the presence of a dislocation between each
pair of columns, with the vectors scaled so that a vector of full
length between the columns corresponds to $1/3$ a Burgers vector.  In
this ground state structure of the dislocation, triads of full length
vectors surrounds the center of each dislocation, corresponding to a
new displacement by a full Burgers vector upon completion of a closed
loop about the each center.  In practice, because of symmetry, the
quadrupole cell may be reduced in half containing two dislocations,
when lattice vectors are properly chosen.

As Section~\ref{sec-bc} notes, for calculations of static properties
such as the ground-state dislocation core structure, the strain fields
from the surrounding dislocations in a quadrupolar array essentially
cancel at each dislocation core.  However in a dynamical problem such
as the calculation of the Peierls stress, the dislocations begin to
interact with the stress fields of the others as they begin to move.
Our previous work\cite{segall} shows that, for the particular geometry
considered here, accurate values for the Peierls stress can be
extracted from quite small unit cells provided the proper procedure is
followed.  We now outline that procedure while reviewing the relevant
background. 

\subsection{Calculation of Peierls stress within periodic boundary conditions } \label{sec:PBCproc}

To calculate the Peierls stress in a small periodic cell, we begin
with lattice vectors appropriate to bulk material in
the absence of dislocations and then, while relaxing the internal
coordinates of the cell, apply increasing pure $\epsilon_{xz}$ strains
(same Cartesian coordinates from Section~\ref{sec:PBCs}) until the
dislocations move.  Such a strain drives the dislocation along the
$[ 11\bar 2 ] $ direction\cite{hirth}.  Before the
dislocation moves, the strain energy of the cell increases
quadratically,
\begin{equation} E =
\frac{1}{2} C' \epsilon_{xz}^2, \label{eqn:energy}
\end{equation}
where $C'$ is an elastic constant associated with the quadrupole unit
cell which can be extracted simply from the energies of the cell as
the strain increases.  The stress associated with this strain is
\begin{equation}
\sigma_{xz} = C' \epsilon_{xz}, \label{eqn:stress}
\end{equation}
so that at the strain at which the dislocation moves,
Equation~(\ref{eqn:stress}) gives the Peierls stress.

The great benefit of the above procedure is that it requires a {\em
minimal} search through phase space in order to calculate the Peierls
stress and accounts accurately for the effects of the
dislocation-dislocation interactions.  The $C'$ elastic constant,
which comes as a direct byproduct of the procedure and requires no
further calculations, suffices to account accurately for the leading
order effects of the inter-dislocation interactions.  This correction
captures most of the effects from working with small unit cells and
can differ from that of an equivalent bulk cell by a value of two or
more.

The above procedure involves several approximations requiring
justification.  First, we apply strain relative to the lattice vectors
of the bulk cell in the absence of the dislocation, rather than those
of the quadrupole array.  Previously, we have shown through explicit
calculation on model inter-atomic potentials that working with the
relaxed lattice vectors of the quadrupolar array does not improve the
value calculated for the Peierls stress and therefore is not
needed\cite{segall}.  The reason for this is that although working
with the bulk lattice vectors generates artificial stresses, these are
primarily diagonal ($\sigma_{ii}$), because the greatest effect of the
presence of the dislocations is to dilate the system\cite{segall}.  In
the present geometry, such diagonal stresses result only in a constant
shift in the energy over the range of applied strain and do not
generate driving (Peach-Kohler\cite{hirth}) forces on the
\mbox{dislocations.}

Some slight care must be taken with the above argument.  Duesbery and
others~\cite{duesbery1,ito} have pointed out that stresses which do
not result in driving forces on a dislocation still may affect the
overall value of the Peierls stress needed to drive the dislocation,
because such stresses may modify the dislocation core
structure~\cite{duesbery2}, an effect not accounted in linear
elasticity theory.  This effect of non-driving stresses is, in fact,
one of the common violations of the Schmid law which bcc metals
exhibit.  Diagonal stresses, however, do not have a large
effect~\cite{duesbery1} on the value for the Peierls stress, as they
tend to only compress or expand the core.  Fortunately, because the
bulk lattice vectors should be relatively close to the that of the
dislocation cell, we expect all of these effects to be quite small, as
we have found previously\cite{segall} and again verify in the test
calculations below.

The second simplification in our procedure is that rather than
applying a stain which imposes a pure $\sigma_{xz}$ stress, we apply a
pure $\epsilon_{xz}$ stain, which also generates a residual
$\sigma_{xy}$ stress\cite{hirth}.  To generate a pure stress of the
form $\sigma_{xz}$, one would have to apply an additional
$\epsilon_{xy}$ stain of a magnitude determined by yet another elastic
constant of the quadrupolar array.  Because the calculation of this
constant would increase significantly the number of calculations required
and because the residual $\sigma_{xy}$ stress\cite{hirth} acts on the
plane perpendicular to the dislocation, and thus does not create a
driving force on the dislocations, we simply apply the pure
$\epsilon_{xz}$ strain.  As with the diagonal stress components,
although the residual in-plane $\sigma_{xy}$ stress does not drive the
dislocations, it can affect the Peierls stress by modifying the core
structure.  Unlike the diagonal stress components, the in-plane stress
does significantly affect the Peierls stress in bcc
metals\cite{duesbery1,ito}.  This effect, however, will be small so
long as the residual $\sigma_{xy}$ stress is small compared to the
driving $\sigma_{xz}$ stress.  The ratio $\sigma_{xy}/\sigma_{xz}$, is
equal to $C''/C'$ where $C'$ is the elastic constant appearing in
Equation~(\ref{eqn:energy}) and $C''$ is another combination of
elastic constants.   In pure bulk cubic materials these constants have
the form\cite{hirth}
\begin{eqnarray*}
C' & = & \frac{1}{3}(C_{11} + C_{44} - C_{12}) \\
C'' & = & -\frac{\sqrt{2}}{6}(2C_{44} + C_{12} - C_{11}),
\end{eqnarray*}
where the $C_{ij}$ are the standard elastic constants for cubic materials.

As evidence of the correlation between the ratio of these constants
and the errors in the Peierls stress, we note that in our previous
study\cite{segall} with same empirical potential as in
Section~\ref{sec-tempdep}, the ratio $C''/C'$ varied from
approximately $1/5$ in the smallest cell studied to less than $1/10$
for all other cells while the error in extracting the Peierls stress
went from 18\% to less than 2\%, respectively.  For the potential used
in that study, the value of $C''/C'$ computed from bulk elastic
constants is $1/90$.  As the bulk value of $C''/C'$ within density
functional theory is less than $1/100$\cite{sohrab_disloc}, we expect
the errors in our {\em ab initio} value of the Peierls stress to be
even somewhat smaller.  Lending further support to this view is the
result of Duesbery and Vitek~\cite{duesbery2} demonstrating that the
effect of $\sigma_{xy}$ stresses on core structure is much less for
the non-degenerate core structure, which we have in our density
functional calculations, than for the degenerate core structures,
which we had in our inter-atomic potential calculations~\cite{segall}.

\subsection{Demonstration}

To demonstrate the efficacy of the above procedure, we now proceed to
extract the Peierls stress in the $\langle 112 \rangle$-direction for
vanadium and tantalum from calculations in small periodic cells, when
the maximum resolved shear stress is along a \{110\}-plane.  In all
calculations we calculate the Peierls stress for an infinite straight
dislocation as our periodic cells has lattice vector $\vec a_3$ =
$a[111]/2$ along the dislocation line.  To allow comparison with the
Peierls stress of isolated dislocations, we employ empirical
potentials for this demonstration.  For vanadium we use the
Finnis-Sinclair~\cite{finnis} potential with modifications made
Ackland and Thetford~\cite{ackland}, and for tantalum we use the same
potential as in Section~\ref{sec-tempdep} but with a slight adjustment
of parameters to produce a non-degenerate core structure.  The bulk
ratios for $C''/C'$ for vanadium and tantalum within these models are
$1/10$ and $1/6.5$, respectively, much larger than the density
functional theory value.

To determine the reference value for the Peierls stress, we employed
cylindrical boundary conditions with large amounts of material,
increasing the radius of the cylinders until the boundary forces were
small~\cite{shenoy} and the Peierls stress approached an asymptotic
value.  To extract the Peierls stress from within periodic boundary
conditions, we follow precisely the procedure which
Section~\ref{sec:PBCproc} outlines.

\begin{figure}
\begin{center}
\rotatebox{90}{\hspace*{0.3in} Energy per Volume eV/\AA$^3$}
\scalebox{0.35}{\includegraphics{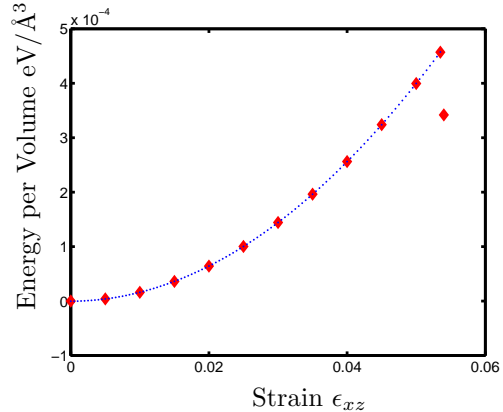}} \\
\hspace*{.5in} Strain $\epsilon_{xz} $
\end{center}
\caption{Energy versus strain of vanadium in a quadrupolar cell of size
42\AA $\times$ 20\AA. Diamonds denote the energy calculated at various
strains. The line is a quadratic fit up to the Peierls stress. }
\label{fig:energy}
\end{figure}

Figure~\ref{fig:energy} shows the resulting energy versus stain curve
for vanadium within a periodic cell of size 42\AA $\times$ 20\AA.  At
a strain $\approx 0.054$, the curve exhibits a discontinuity signaling
the critical strain for moving the dislocation.  The curvature of the
fit determines the elastic constant $C'$ through
Equation~(\ref{eqn:energy}).  Finally, combining this value of $C'$
with the observed critical strain Equation~(\ref{eqn:stress}) yields
the Peierls stress.  We repeated this procedure for tantalum as well.

Table~\ref{table:energyvscell} summarizes our results for both
vanadium and tantalum.  The table shows that the errors are relatively
small, much smaller than the general discrepancy between empirical
potentials and the experimental extrapolations.  It is also noteworthy
that the potentials employed in this demonstration exhibit $C''/C'$
ratios over an order of magnitude larger than density functional
theory.  From these and previous results~\cite{segall}, we
conservatively estimate that the error in the Peierls stress in the
density functional calculations below should be no greater than
$\approx 20\%$.

\begin{table}
\centering
\begin{tabular}{|l|c|c|}
\hline  & 23\AA $\times$ 11.5\AA  & 42\AA $\times$ 20\AA \\
\hline Ta   & 25\% & 10\% \\
\hline V   &  26\% & 11\% \\
\hline
\end{tabular}
\caption{Magnitude of percentage error in calculating the Peierls
stress in periodic cells of two different sizes from empirical
potentials for vanadium and tantalum.}
\label{table:energyvscell}
\end{table}

\section{Density functional results and discussion} \label{sec-results}

\subsection{Computational Details}

All of our first principles electronic structure calculations employ
the plane-wave density-functional pseudopotential
approach~\cite{payne} within the local density
approximation~\cite{ceperley,perdew}.  We employ pseudopotential of
the Kleinman-Bylander form~\cite{kleinman} with $s$, $p$ and $d$
non-local channels which has been used successfully in previous
works\cite{woodward_ps,sohrab_disloc} and a plane wave basis with
cutoff of 40~Rydberg.  As justified above, we employ a super-cell
containing a quadrupolar array of dislocations of size $\vec r_1 =
5a[1,\bar 1,0], \ \vec r_2 = (3/2)a[1,1,\bar 2], \ \mbox{and} \ \vec
r_3 = a[1,1,1]/2$, where $a=3.25$~\AA \ is the lattice constant of the
cubic unit cell. The lattice vectors of this cell are $\vec a_1 = \vec
r_1/2 - \vec r_2 + \vec r_3/2, \ \vec a_2 = \vec r_1/2 + \vec r_2 +
\vec r_3/2 \mbox{ and } \vec a_3 = \vec r_3$. To carry out the
integrations over the Brillouin zone we use a non-zero electronic
temperature of $k_BT=0.1$~eV to facilitate integration over the Fermi
surface and sample the zone at sixteen special $k$-points~\cite{kpts}.
These choices give energy differences reliably to within 0.1 eV/atom.
Finally, to determine the electronic structure, we minimize using the
analytically continued functional approach~\cite{arias}, expressed
within the DFT++~\cite{dftpp} formalism.

\begin{figure}
\begin{center}
 Ta \hspace*{1.3in} Mo \\
\rotatebox{90}{ \hspace*{.3in} DFT}
\scalebox{0.325}{\includegraphics{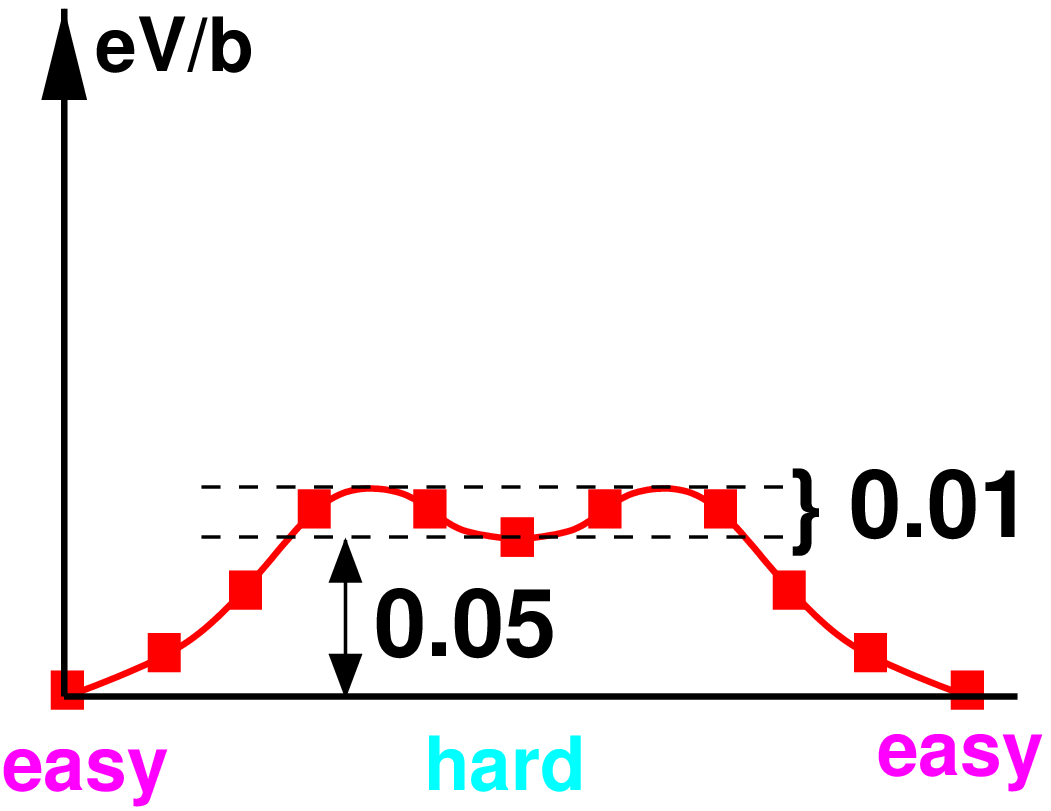}}  \
\scalebox{0.325}{\includegraphics{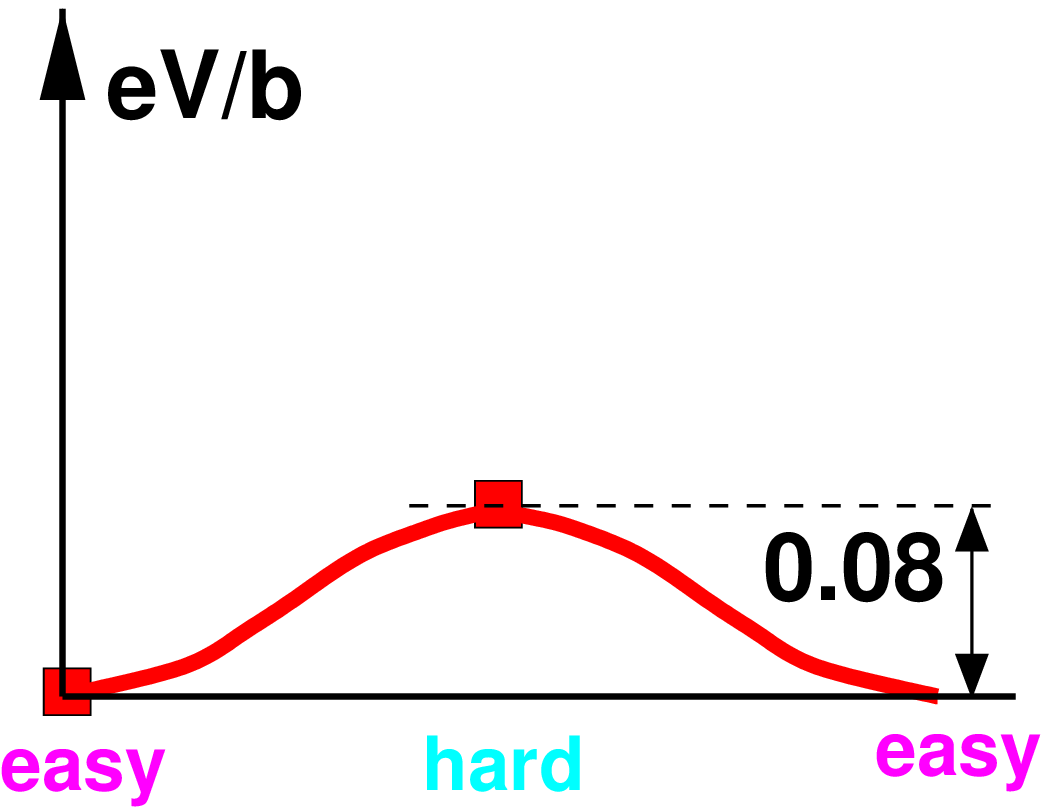}}  \\
\rotatebox{90}{ \hspace*{.25in} MGPT}
\scalebox{0.325}{\includegraphics{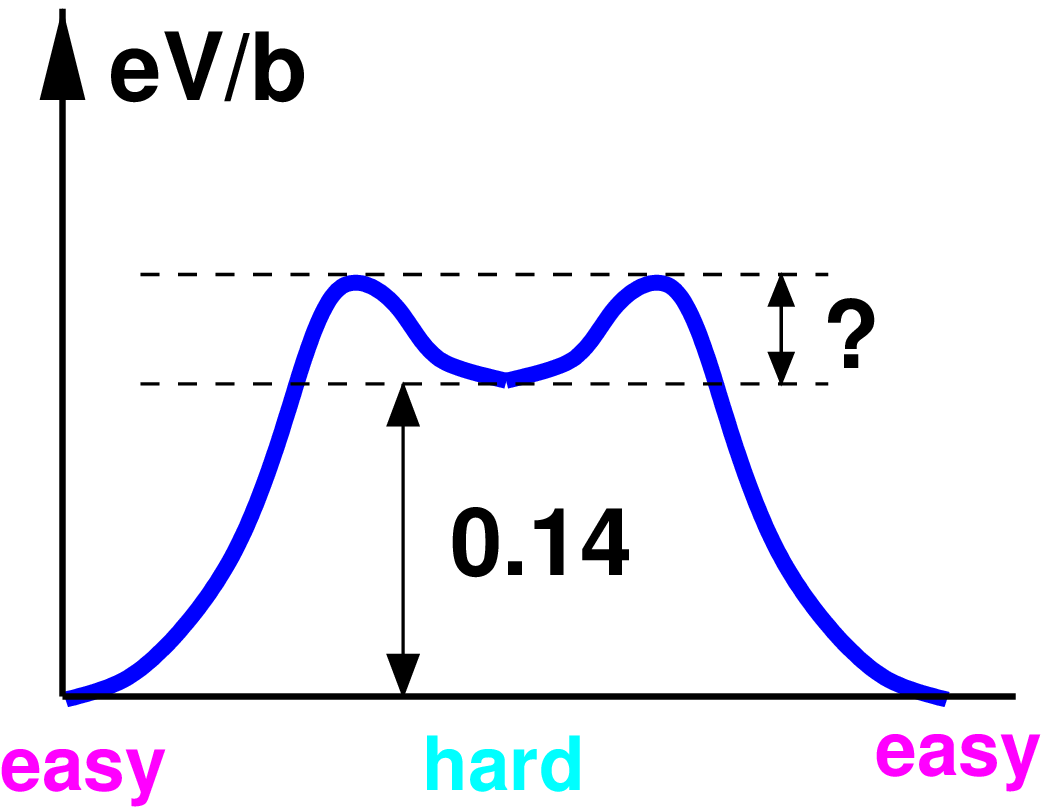}}  \
\scalebox{0.325}{\includegraphics{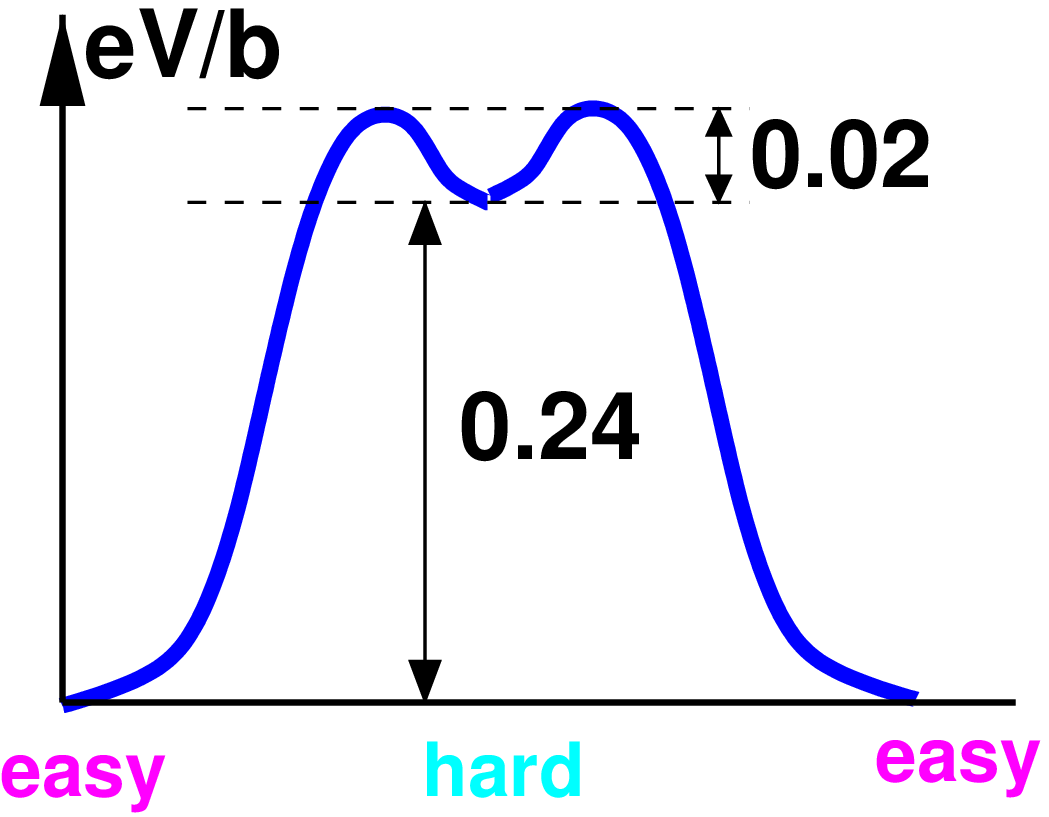}}
\end{center}
\caption{Energy landscape per Burgers vector along a reaction
coordinate when going from the easy core to the hard core
configurations in tantalum and molybdenum: density functional theory
(squares in upper panels), MGPT (curves in lower panels).  {\em Ab
initio} results for molybdenum are from
Reference~\onlinecite{sohrab_disloc}, and the MGPT results for tantalum are
from Reference~\onlinecite{moriarty}.}
\label{fig:mgptvsdft}
\end{figure}

\subsection{Energy landscapes}

To demonstrate the discrepancies that occurs between the predictions
of empirical potentials and first principles electronic structure
studies, we compare the energy landscape for a dislocation moving
along a reaction coordinate from the easy core configuration to the
hard core configuration\cite{vitek,xu} for both molybdenum and
tantalum as determined within MGPT\cite{mgpt} and as calculated {\em
ab initio}.  Within MGPT, we have carried out the calculation for
molybdenum ourselves and we used the hard-easy core energy difference
reported in the literature\cite{moriarty} for tantalum.  Within density
functional theory, we have calculated the energy at a number of points
along the reaction pathway for tantalum, and for molybdenum we report
the difference between the easy core and hard core configurations as
found in Reference~\onlinecite{sohrab_disloc}. We also note that, for
molybdenum, within density functional theory, the hard core
configuration was not stable and therefore the stable structure found
within MGPT was used as the reference state.

Figure~\ref{fig:mgptvsdft} shows the results.  Most noticeably, the
atomistic landscapes are three times stiffer than the {\em ab initio}
landscapes.  This raises the question whether the approximate factor
of three overestimate of theoretical calculations over the
extrapolation of the experimental Peierls stresses to zero-temperature
is due to defects in the inter-atomic potentials or to failures in the
connection between the experiments and the theoretical calculations.

\subsection{Verification of cell size}

\begin{figure}
\begin{center}
\rotatebox{90}{\hspace*{0.7in} {\Large $[  1 1 \bar 2 ]  \rightarrow$ }}
\scalebox{0.4}{  \ \ \ \ \ \includegraphics{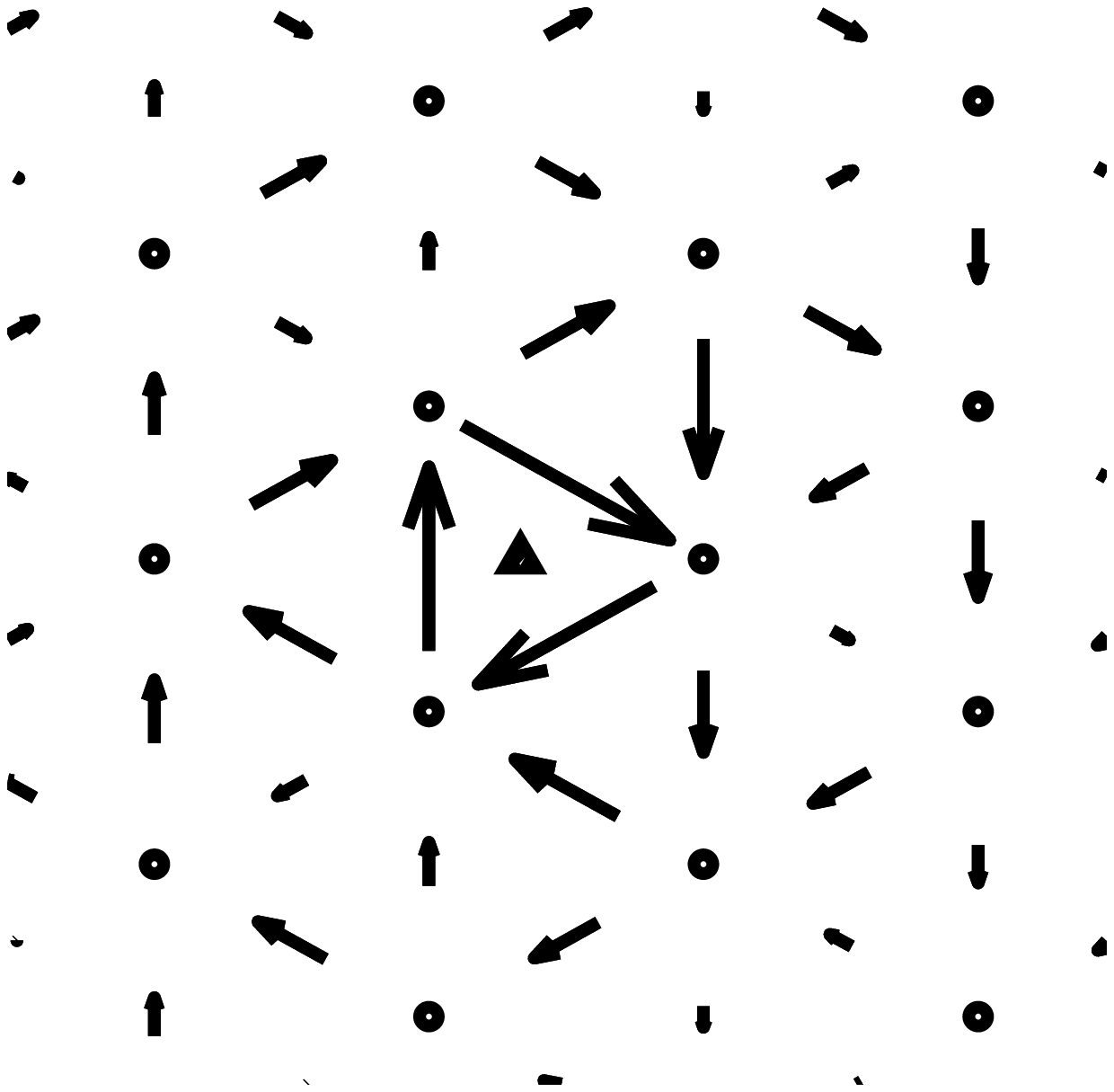}} \\ \ \\
\hspace*{0.4in} {\Large $[  1 \bar 1 0 ] \rightarrow$}
\end{center}
\caption{Easy core structure for tantalum calculated within density
functional theory. This cell of size 41\AA $\times$ 20 \AA \ gives
very similar results to the cell of size 23\AA $\times$ 12\AA, used in
Reference~\onlinecite{sohrab_disloc}. The solid triangle in the figure
represents the center of the dislocation.}
\label{fig:easy135core}
\end{figure}

To compute the Peierls stress, we shall employ a cell of dimensions
23\AA $\times$ 12\AA.  To verify that long-range electronic structure
effects in metals do not interfere with results in such a cell, we
compare the core structure reported previously\cite{sohrab_disloc} for
this cell with a new calculation using a larger cell.
Figure~\ref{fig:easy135core} shows the result for the core structure
in cell of size 41\AA $\times$ 20\AA. The quadrupole array has size
$\vec r_1 \ = \ 9a[1,\bar 1,0], \ \vec r_2 \ = \ (5/2)a[1,1,\bar 2], \
\mbox{and} \ \vec r_3 \ = \ a[1,1,1]/2$. The lattice vectors are $\vec
a_1 = \vec r_1/2 - \vec r_2 + \vec r_3/2, \ \vec a_2 = \vec r_1/2 +
\vec r_2 + \vec r_3/2 \mbox{ and } \vec a_3 = \vec r_3$. Here we use
eight special $k$-points~\cite{kpts2}, which is sufficient as the
lattice vectors in the plane of the dislocation have doubled.

We note that the core structure is very similar to previous
results\cite{sohrab_disloc} which used a smaller cell equal in size to
that we shall employ for our calculation of the Peierls stress.  We
therefore do not expect the long-range nature of electronic effects in
metallic systems to greatly affect the value which shall extract for
the Peierls stress.  We also note that the empirical potential results
for tantalum (Section~\ref{sec-conv}) had a very large cutoff of 9\AA
\ and accurate results were obtained in the smallest cells used in
those calculations.  These facts lend confidence in the reliability of
our density-functional theory predictions below.

\subsection{Results for the Peierls Stress}

\begin{figure}
\begin{center}
\rotatebox{90}{\hspace*{.2in} Energy per Volume [eV/\AA$^3$]}
\scalebox{0.4}{\includegraphics{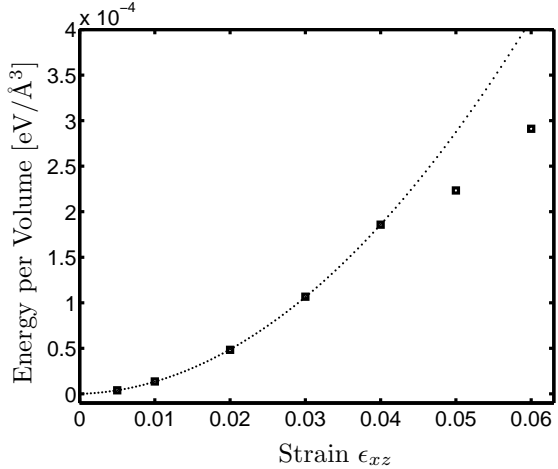}} \\
\hspace*{0.5in} Strain $\epsilon_{xz} $
\end{center}
\caption{Energy versus strain for tantalum in a cell of size 23\AA
$\times$ 12\AA, calculated within density functional theory. Squares
denote the calculate energy at various strains. The line is a
quadratic fit to the first five data points.  The graph indicates that
the points with strain at and above 0.05 ($\sigma_{xz}$ = 1.76~GPa)
will be at a stress above the Peierls stress. }
\label{fig:energydft}
\end{figure}

Figure~\ref{fig:energydft} shows our density functional theory results
for energy as a function of strain following the procedure of
Section~\ref{sec:PBCproc}. We regard each data point as fully relaxed
when the magnitude of the residual force on each atom is less than
0.005~eV/\AA.

\begin{figure}
\begin{center}
\rotatebox{90}{\hspace*{0.5in} {\Large $[ 1 1 \bar 2 ] \rightarrow$ }}
\ \ \ \scalebox{0.4}{\includegraphics{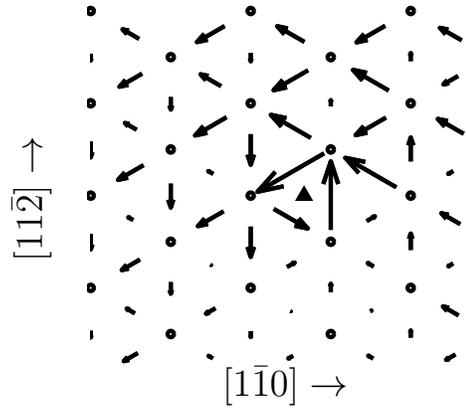}} \\
\hspace*{0.4in} {\Large $[ 1 \bar 1 0 ] \rightarrow$}
\end{center}
\caption{Differential displacement map at an applied strain of
  0.04. ($\sigma_{xz}$ = 1.41~GPa) The cell has converged to within
  the tolerance defined in the text. The solid triangle in the figure
  is the location of the center of the dislocation, under no
  stress. The center of the dislocation, at this applied stress, has
  not moved.}
\label{fig:dd8}
\end{figure}

\begin{figure}
\begin{center}
\rotatebox{90}{\hspace*{0.5in} {\Large $[ 1 1 \bar 2 ]
\rightarrow$ }} \ \ \ \scalebox{0.4}{\includegraphics{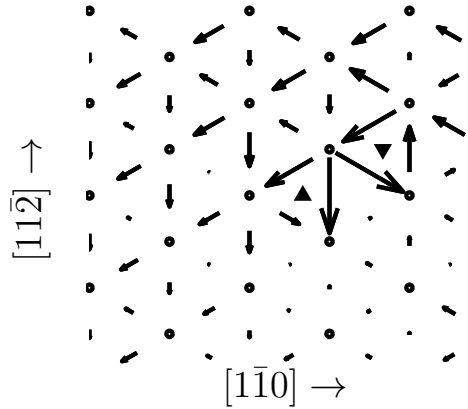}}
\\
\hspace*{0.4in} {\Large $[ 1 \bar 1 0 ] \rightarrow$}
\end{center}
\caption{Differential displacement map at an applied strains of
  0.05. ($\sigma_{xz}$ = 1.76~Gpa) The cell has converged to within
  the tolerance defined in the text. This figure shows that the core
  has moved from the old triad (solid triangle) to a new triad
  (upside-down, solid triangle) along a $\langle
  112\rangle$-direction.}
\label{fig:dd10}
\end{figure}

From Figure~\ref{fig:energydft} it is apparent that, at a strain of
0.05 the Peierls stress has been exceeded, whereas at a strain of 0.04
the energy curve lies on the elastic solution. From the curvature of
the data in the elastic region ($C'$ = 35.2 GPa) and these critical
strains, we bound the Peierls stress ($\sigma_{xz}$) to be between
1.41~GPa and 1.76~GPa.  These results are slightly lower than the
previous density functional results of $\approx$ 1.8~GPa obtained
using Greens function boundary conditions\cite{woodward2}.  However
both results are still in reasonable agreement, as we expect our
results to be within 20\% of the infinite cell limit, while it is
unclear how the domain boundary affected the value of the Peierls
stress in the Greens function boundary condition calculation.

Figures~\ref{fig:dd8}~and~\ref{fig:dd10} show differential
displacement maps\cite{vitek} of the dislocation configuration at
strains of 0.04 and 0.05, respectively. The solid triangles in the
figures represent the location of the center of the dislocation before
the application of stain and the upside-down, solid triangle represents
the center of the dislocation once it has moved.  Figure~\ref{fig:dd8}
shows the relaxed dislocation core to remain in the position of its
unstressed state, whereas Figure~\ref{fig:dd10} shows that at an
applied strain of 0.05, the dislocation moves onto the next triad of
columns of atoms. This glide is consistent with such screw
dislocations as the initial displacement is along the (01$\bar
1$)-plane. The subsequent motion should be along the (1$\bar
1$0)-plane, and hence overall motion is along a \{112\}-plane
(twinning direction), consistent with the previous density functional
theory~\cite{woodward2} calculations.

Intriguingly, despite the fact that the {\em ab initio} energy
landscape is {\em less} corrugated than that of the inter-atomic
potential by a factor of nearly three (Figure~\ref{fig:mgptvsdft}),
the above {\em ab initio} result for the Peierls stress is over a
factor of two {\em larger} than the empirical potential result
(Table~\ref{table_peierls_orint}).  Moreover, our result is over a
factor of five {\em larger} than extrapolations of experimental data
to zero-temperature\cite{tang}, a discrepancy much larger than any
effects from the use of either our relatively small quadrupolar cell
or the local density approximation to density-functional
theory. Certainly, as Section~\ref{sec-tempdep} shows, some of this
discrepancy can be due to the extrapolation of experimental data to
zero temperature.  However, errors in the zero-temperature
extrapolation may not likely account for all of the discrepancy as the
underestimation in Figure~\ref{fig:tempdep} is only $\approx 10-20\%$.
One must therefore consider the possibility of other factors such as
the effects of kinks, defects, surfaces, strain-rate dependencies and
screening from the presence of other dislocations, to accurately
predict the experimental results.

\section{Conclusion}
This work explores various aspects of the the Peierls stress for the
$\langle 111 \rangle$ screw dislocation in bcc tantalum.  The first
non-zero temperature results for the Peierls stress in this system
shows both a strong orientation and temperature dependent response,
consistent with experimental results.  This data also demonstrate
that common extrapolations of experimental data
tend to underestimate the zero-temperature limit.

We have also presented the first density functional theory calculation
for the Peierls stress within perioidic boundary conditions, the
approach best suited to metallic systems. The value we find for the
Peierls stress is substantially larger than both the experimental
extrapolations and current empirical potential results.  This
difference is much larger than errors which could normally be
attributed to the use of a relatively small unit cell or the local
density approximation to density-functional theory.  The error is also
significantly larger than the underestimation we have seen in the the
extrapolation of nonzero-temperature data to zero temperature, thus
supporting the notion that mechanisms other than the simple Peierls
resistance may play an important role in controlling the process of
low temperature slip.

\section{Acknowledgments}
The authors would like to thank Guofeng Wang for providing us with the
new parameters for qEAM potential. This work was supported by an ASCI
ASAP Level 2 grant (Contracts No. B338297 and
No. B347887). Computational support on ASCI Blue-Pacific provided
through the Cal-Tech DOE ASCI center.  We thank the members of the
H-division at Lawrence Livermore National Laboratories for providing
the Ta pseudopotential, the Mo MGPT code, and many useful discussions.

\bibliographystyle{unsrt}

\begin{thebibliography}{10}
\bibitem{kpts} The $k$-point set is: $\pm\frac{1}{32}$(\{1,1,2\},
 \{17,-15,2\}, \{3,3,6\}, \{19,-13,6\}, \{5,5,10\}, \{21,-11,10\},
 \{7,7,14\}, \{23,-9,14\}). Note, that because of symmetry our
 quadrupole cell has only two dislocations.  Without taking advantage
 of this symmetry the cell would double and have four dislocation with
 lattice vectors $\vec b_1=5a[1,\bar 1,0], \vec b_2 = 3a[1,1,\bar 2]$
 and $\vec b_3 = a[1,1,1]/2$. The above $k$-points, for the smaller
 cell, would then map onto eight special $k$-points: $k_1 = k_2 = 0, \
 k_3 \in \pm \{
 \frac{1}{16},\frac{3}{16},\frac{5}{16},\frac{7}{16}\}$, for the
 larger cell, where $k_1, k_2$ and $k_3$ are generated from the
 reciprocal lattice of $\{\vec b_i\}$.
\bibitem{kpts2}
In the larger cell the  $k$-point set is: $\pm\frac{1}{32}$(\{1,1,2\},
 \{3,3,6\}, \{5,5,10\}, \{7,7,14\}).
\end{thebibliography}

\end{document}